\documentclass[twocolumn,superscriptaddress]{revtex4-1}
\usepackage{graphicx}
\usepackage{epstopdf}
\usepackage{upgreek}
\usepackage{amsmath}
\usepackage{amssymb}
\usepackage{color}
\usepackage{soul}
\usepackage{ulem}
\usepackage{float}
\usepackage[utf8]{inputenc}
\usepackage[colorlinks,linkcolor=blue,
            citecolor=blue,
            hyperindex,
            pdfstartview=FitH,
            plainpages=false]
            {hyperref}

\begin{document}
\title{Spin-Valley Locking in 2$\it{H}$-TaS$_2$ and Its Co-Intercalated Counterpart: Roles of Surface Domains and Co Intercalation}

\author{Hai-Lan Luo} \affiliation{Department of Physics, University of California at Berkeley, Berkeley, CA 94720, USA} \affiliation{Materials Sciences Division, Lawrence Berkeley National Laboratory, Berkeley, CA 94720, USA}
\author{Josue Rodriguez} \affiliation{Department of Physics, University of California at Berkeley, Berkeley, CA 94720, USA}
\author{Maximilian Huber} \affiliation{Materials Sciences Division, Lawrence Berkeley National Laboratory, Berkeley, CA 94720, USA}
\author{Haoyue Jiang} \affiliation{Materials Sciences Division, Lawrence Berkeley National Laboratory, Berkeley, CA 94720, USA}\affiliation{Graduate Group in Applied Science and Technology, University of California at Berkeley, Berkeley, CA 94720, USA}
\author{Luca Moreschini} \affiliation{Materials Sciences Division, Lawrence Berkeley National Laboratory, Berkeley, CA 94720, USA}\affiliation{Department of Physics, University of California at Berkeley, Berkeley, CA 94720, USA}
\author{Pranav Thekke Madathil}\affiliation{Department of Physics, University of California at Berkeley, Berkeley, CA 94720, USA}
\author{Catherine Xu} \affiliation{Department of Physics, University of California at Berkeley, Berkeley, CA 94720, USA}
\author{Chris Jozwiak} \affiliation{Advanced Light Source, Lawrence Berkeley National Laboratory, Berkeley, CA 94720, USA}
\author{Aaron Bostwick} \affiliation{Advanced Light Source, Lawrence Berkeley National Laboratory, Berkeley, CA 94720, USA}
\author{Alexei Fedorov} \affiliation{Advanced Light Source, Lawrence Berkeley National Laboratory, Berkeley, CA 94720, USA}
\author{James G. Analytis} \affiliation{Department of Physics, University of California at Berkeley, Berkeley, CA 94720, USA}\affiliation{CIFAR Quantum Materials, CIFAR, Ontario, Toronto M5G 1M1, Canada}\affiliation{Kavli Energy NanoScience Institute, University of California, Berkeley, CA 94720, USA}
\author{Dung-Hai Lee} \affiliation{Department of Physics, University of California at Berkeley, Berkeley, CA 94720, USA}\affiliation{Materials Sciences Division, Lawrence Berkeley National Laboratory, Berkeley, CA 94720, USA}
\author{Alessandra Lanzara} \email{alanzara@lbl.gov} \affiliation{Department of Physics, University of California at Berkeley, Berkeley, CA 94720, USA} \affiliation{Materials Sciences Division, Lawrence Berkeley National Laboratory, Berkeley, CA 94720, USA}\affiliation{Kavli Energy NanoScience Institute, University of California, Berkeley, CA 94720, USA}

\date{\today}

\begin{abstract}
\noindent {\bf Abstract:} Tuning and probing spin-valley coupling is key to understanding correlated ground states in 2$\it{H}$-TaS$_2$. Its magnetically intercalated analogue, Co$_{1/3}$TaS$_2$, introduces additional degrees of freedom, including modified interlayer coupling and magnetism, to modulate spin-valley physics. Surface-sensitive probes like ARPES are essential for accessing surface spin texture, yet previous studies on 2$\it{H}$-TMDs have reported conflicting results regarding spin-polarized bands, leaving open whether these discrepancies are intrinsic or extrinsic. Here we performed spatially resolved spin-ARPES measurements on 2$\it{H}$-TaS$_2$ and Co$_{1/3}$TaS$_2$. Our results reveal robust spin-valley locking on both compounds. Importantly, Co intercalation enhances interlayer hybridization and introduces magnetism while preserving the TaS$_2$-derived spin texture. We further observe a spatial reversal of the out-of-plane spin polarization, which we attribute to different surface domains. This effect complicates quantifying spin textures and may underlie prior inconsistent observations. Our findings provide microscopic insight into how interlayer interactions and surface domains together govern spin-valley phenomena in layered TMDs.

\vspace{3mm}
Keywords: Spin-valley locking, Transition metal dichalcogenides, Spin-resolved angle-resolved photoemission spectroscopy (ARPES), surface domains
\end{abstract}

\maketitle

\vspace{6mm}

In layered transition metal dichalcogenides (TMDs), metallic 2$\it{H}$-TaS$_2$ is particularly notable for exhibiting both unconventional Ising superconductivity and charge density wave (CDW) order\,\cite{SNagata1992KTsutsumi,IGuillamon2011ECoronado,EMoratalla2016ECoronado,YYang2018PHerrero}. In an atomic layer, inversion asymmetry combined with strong spin-orbit coupling (SOC) gives rise to valley-dependent spin splitting, namely spin-valley locking\,\cite{ZZhu2011UShwin,DXiao2012WYao}, which is believed to stabilize the Ising superconducting phase\,\cite{SBarrera2018BHunt,JLu2015JYe} and to shape the momentum-dependent CDW gap\,\cite{DRahn2012KRossnagel}. Crucially, spin-valley locking in TMDs can be tuned through various mechanisms, including modulation of interlayer coupling (e.g., via chemical intercalation\,\cite{Zli2024JLu,ZWan2024XDuan} or applied pressure\,\cite{DFreitas2016HSuderow}), magnetic proximity coupling\,\cite{TNorden2019HZeng}, and the application of external magnetic fields\,\cite{HYuan2013YIwasa}. Consequently, tuning and probing the spin texture in 2$\it{H}$-TaS$_2$ is essential for understanding its complex correlated ground states.

Among these tuning approaches, magnetic intercalation is particularly intriguing. Magnetically intercalated TMDs host diverse magnetic structures and electronic properties\,\cite{SParkin1980RFriend,SParkin1980RFriend2,NNair2020JAnalytis,KLu2020GMacDougall,HTakagi2023SSeki,LXie2023DBediako,BEdwards2023PKing,ZMeng2025YHou,PGu2025YYe}, and by modulating the interlayer coupling while simultaneously introducing long-range magnetic order, they provide an an extra degree of freedom for controlling spin-valley physics. We focus here on Co$_{1/3}$TaS$_2$, where cobalt (Co) atoms occupy the van der Waals gaps between TaS$_2$ layers, exhibiting a long-range magnetic order\,\cite{HTakagi2023SSeki,PPark2023JPark,PPark2024JPark}, changing the band filling\,\cite{HLuo2025ALanzara} and potentially mediating the interlayer hopping. Such a magnetically intercalated system provides a unique platform to explore how interlayer interactions, magnetism, and spin-valley coupling are intertwined\,\cite{BEdwards2023PKing}.

While various approaches can tune the spin-valley coupling, direct experimental access to the underlying spin texture in bulk TMDs remains technically demanding and relies on surface-sensitive probes. In AB-stacked bulk 2$\it{H}$-TMDs (whether pristine or intercalated), the spin polarization of one layer is usually compensated by that of its 180$^{\circ}$ rotated neighbouring layer, yielding overall spin-degenerate bulk states as required for a centrosymmetric material. In principle, by using surface-sensitive techniques, such as spin- and angle-resolved photoemission spectroscopy (ARPES), a so-called ``hidden spin polarization" can be observed on the topmost layer of 2$\it{H}$-TMDs, where the local inversion symmetry is broken\,\cite{XZhang2014AZunger,XXu2014THeinz,JRiley2014PKing,LBawden2016PKing,ERazzoli2017PAebi}.

However, prior reports on 2$\it{H}$-TMDs are contradictory: some observe spin-valley locking, whereas others report spin-degenerate states with no detectable polarization\,\cite{RSuzuki2014YIwasa, AAlmoalem2024AKanigel,LBawden2016PKing}. The origin of this discrepancy\,$-$\,whether intrinsic or arising from extrinsic effects like nonequivalent surface terminations\,$-$\,has largely been overlooked. Thus, experimentally resolving the spin texture in both pristine and intercalated 2$\it{H}$-TaS$_2$, particularly at the microscopic scale, is crucial for disentangling the roles of inversion symmetry, interlayer interaction, and magnetic ordering in shaping the spin-valley band topology.

In this work, we use spin-ARPES to directly image the surface spin textures of both 2$\it{H}$-TaS$_2$ and its magnetically intercalated counterpart, Co$_{1/3}$TaS$_2$. Our measurements show clear signatures of spin-valley locking on the surfaces of both compounds, in line with expectations for monolayer 1$\it{H}$-TaS$_2$. We further find that the antiferromagnetic ordering of Co atoms acts as an interlayer bridge, enhancing interlayer interactions without introducing significant perturbations to the top-layer spin texture. However, determining the magnitude of spin polarization is complicated by a domain-dependent inversion of the out-of-plane spin component across the sample surface. Spatially resolved spin-ARPES measurements suggest that this inversion arises from two distinct TaS$_2$ surface terminations (likely corresponding to the A and B layers of the AB-stacking or to different rotational domains). This extrinsic effect introduces substantial uncertainty in interpreting spin textures across different domains and samples and must be carefully accounted for in spin-resolved measurements. Our results provide microscopic insight into how local symmetry breaking, magnetic order, and interlayer coupling collectively shape the spin-valley landscape in layered and intercalated 2$\it{H}$-TaS$_2$.

\vspace{6mm}

\begin{figure*}[!htb]
\begin{center}
\includegraphics[width=2\columnwidth,angle=0]{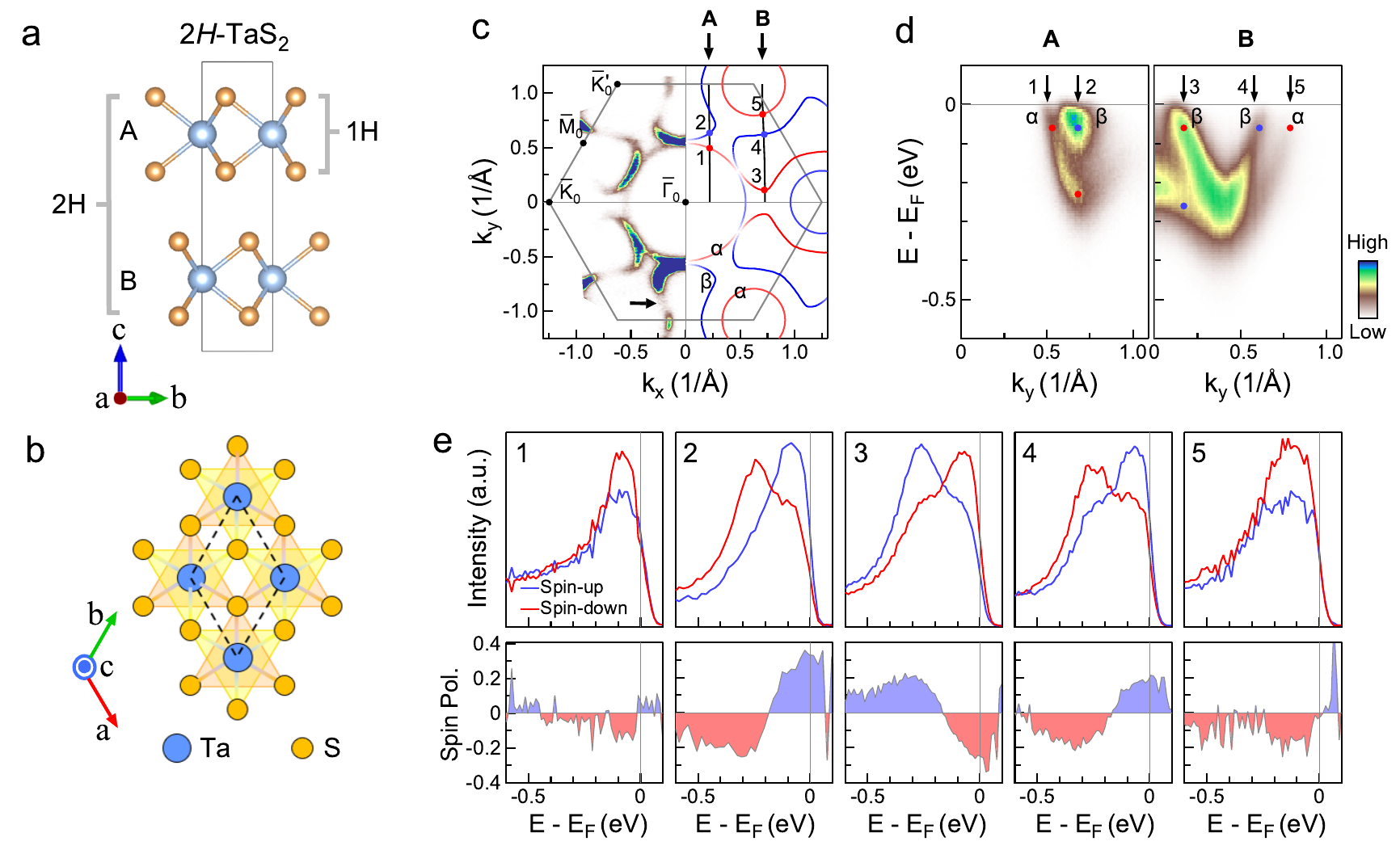}
\end{center}
\caption{\label{fig1} \textbf{Spin-valley locking in 2$\it{H}$-TaS$_2$.} {\textbf{a and b,}} Side view (a) and top view (b) of the crystal structure of 2$\it{H}$-TaS$_2$. {\textbf{c,}} Fermi surface mapping of 2$\it{H}$-TaS$_2$ measured with 55\,eV photons (left), together with the corresponding spin-polarized Fermi surface contours (right). Blue and red denote the spin-up and spin-down components of the out-of-plane spin polarization, respectively. Solid gray lines mark the Brillouin zone of 2$\it{H}$-TaS$_2$. {\textbf{d,}} Spin-integrated band structures of 2$\it{H}$-TaS$_2$ along momentum cuts\,A and B (as marked in c), measured at 55\,eV. {\textbf{e,}} Spin-resolved energy distribution curves (EDCs) (upper panels) and corresponding spin polarizations (lower panels) taken at locations 1-5 marked in (c,\,d).
}
\end{figure*}

\vspace{2mm}

2$\it{H}$-TaS$_2$ crystallizes in the hexagonal $P$6$_3$/mmc space group. Figures~\ref{fig1}a and~\ref{fig1}b show the side and top views of its crystal structure. The unit cell consists of two layers of 1$\it{H}$-TaS$_2$ stacked along the $c$\,-\,axis with 180$^{\circ}$ in-plane ($a$-$b$ plane) rotation. Figure~\ref{fig1}c displays the Fermi surface mapping of bulk 2$\it{H}$-TaS$_2$, with the left panel showing the raw data and the right panel presenting the experimentally extracted Fermi surface contours. Two distinct bands cross the Fermi level: the $\alpha$ band forms two circular hole-like pockets centered at $\bar\Gamma_0$ and $\rm\bar{K}_0$ (the K-valley center), with the $\rm\bar{K}_0$-centered pocket appearing weaker in intensity; the $\beta$ band exhibits a dumbbell-shaped electron-like pocket around $\rm\bar{M}_0$, which is partially gapped due to CDW formation, as indicated by the arrow. ARPES dispersions along the cut A and cut B directions are presented in Fig.~\ref{fig1}d. Spin-resolved ARPES measurements were performed at selected momenta (points 1-5), as marked in Fig.~\ref{fig1}c and \ref{fig1}d. Figure~\ref{fig1}e displays the out-of-plane spin-resolved energy distribution curves (EDCs) (spin-up $I_\uparrow$ and spin-down $I_\downarrow$ components) and the corresponding spin polarization, defined as: $\it{P}$\,=\,$\frac{1}{S}$$\frac{I_\uparrow-I_\downarrow}{I_\uparrow+I_\downarrow}$,
where $S$ is the Sherman function. By comparing spin polarization at different binding energies, we find that the $\alpha$ and $\beta$ bands exhibit opposite spin polarizations up to $\sim$35\% at the same momenta. Additional spin-resolved data (Supplementary Fig.~S1) reveal that the spin polarization of both the $\alpha$ and $\beta$ Fermi surfaces reverses sign upon crossing the $\bar\Gamma_0$--$\rm\bar{M}_0$ line, clearly establishing a spin-valley locking scenario. For visual clarity, the spin textures on the Fermi surface are color coded blue (spin-up) and red (spin-down) in Fig.~\ref{fig1}c.

In contrast, some previous studies on 2$\it{H}$-TMDs reported a complete absence of spin-valley locking\,\cite{RSuzuki2014YIwasa,AAlmoalem2024AKanigel}. To explore the origin of this discrepancy, we consider factors that suppress the observable spin polarization in bulk 2$\it{H}$-TaS$_2$. In monolayer 1$\it{H}$-TMDs, where inversion symmetry is broken, spin-orbit coupling (SOC) produces valley-dependent spin splitting of the $\alpha$ and $\beta$ bands, with electron spins pinned out of plane\,\cite{ZZhu2011UShwin,DXiao2012WYao}. This property has been exploited in valleytronic devices based on monolayer 1$\it{H}$-TMDs\,\cite{HZeng2012XCui,KMak2012THeinz}. In bulk 2$\it{H}$-TMD, two main factors can diminish the measured spin polarization at the top-layer relative to that of an isolated monolayer. First, interlayer coupling can hybridize wavefunctions between adjacent layers with opposite spin polarizations, thereby weakening the net signal. The polarization scales approximately as $P\,\propto\,\frac{1}{\sqrt{1 + \left(t_{\perp}/\Delta_{\rm{SOC}}\right)^2}}$, where $t_{\perp}$ and $\Delta_{\rm{SOC}}$ denote the interlayer hopping and SOC strength, respectively\,\cite{XZhang2014AZunger}. Second, since ARPES probes photoelectrons emitted from a finite depth, the top-layer spin polarization can be partially compensated by the opposite spin orientation from the underlying layer. This gives an additional reduction factor of approximately $\frac{1}{1 + e^{-\Delta L / L}}$, with $\Delta L$ the interlayer spacing and $L$ the photoelectron mean-free path\,\cite{XZhang2014AZunger}. Nevertheless, even with these attenuation effects, the measured spin polarization in bulk 2$\it{H}$-TaS$_2$ is not expected to vanish. Two pieces of evidences support this point: $\textbf{(\romannumeral1)}$ clear spin-valley locking has been observed in 2$\it{H}$-NbSe$_2$\,\cite{LBawden2016PKing}, which has an even larger $t_{\perp}/\Delta_{\rm{SOC}}$ ratio than 2$\it{H}$-TaS$_2$\,\cite{SBarrera2018BHunt}; $\textbf{(\romannumeral2)}$ using a photoelectron mean-free path of $\sim$\,7\,\AA~(corresponding to the photon energy used here), we estimate that the measured spin polarization retains about 78\% of the top-layer value. Thus, the spin-valley locking observed in our experiments likely reflect intrinsic properties of 2$\it{H}$-TaS$_2$. Possible reasons why some previous studies did not detect spin polarization in 2$\it{H}$-TaS$_2$ are discussed below.
\vspace{3mm}


\begin{figure*}[!htb]
\begin{center}
\includegraphics[width=2\columnwidth,angle=0]{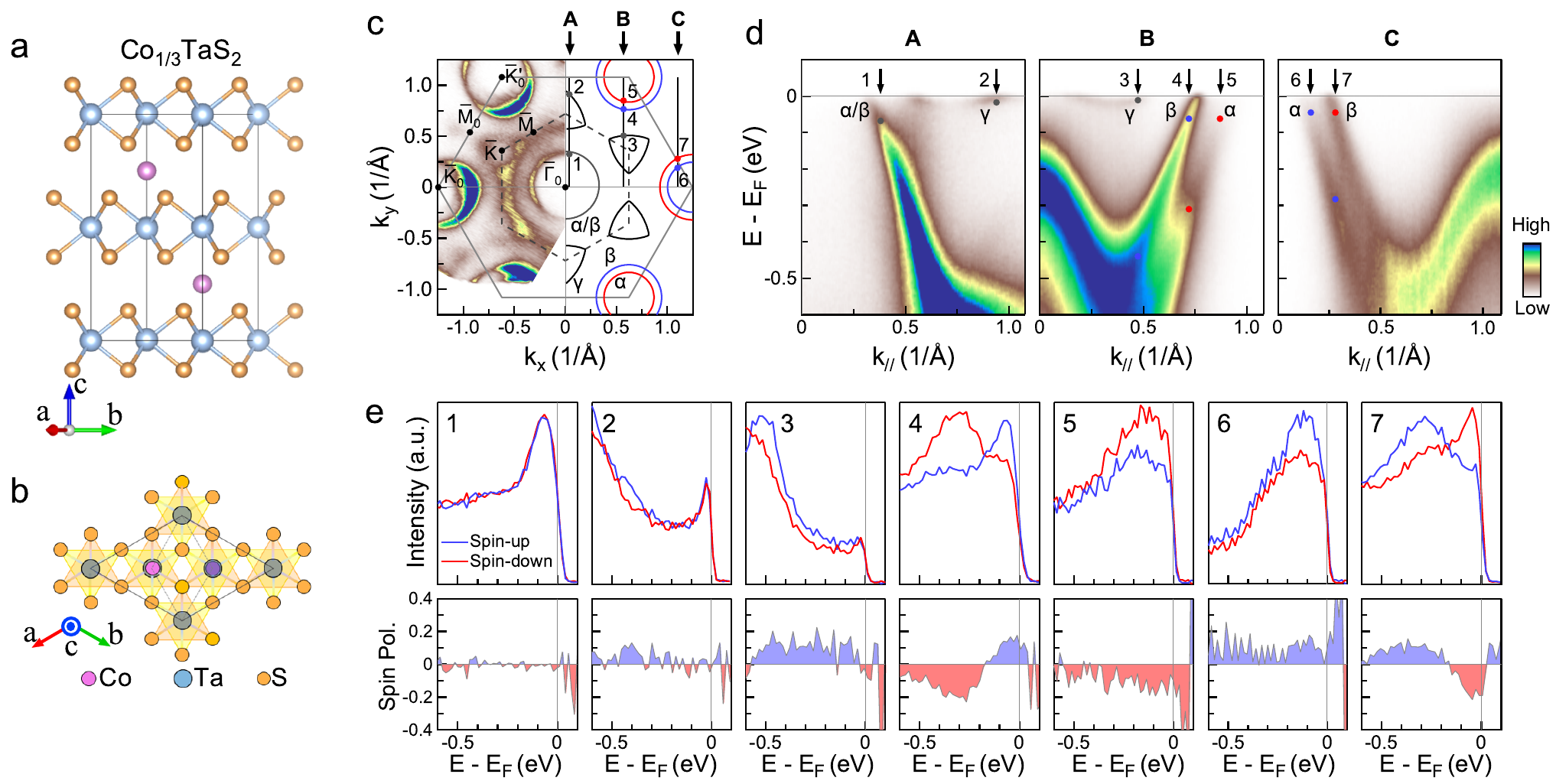}
\end{center}
\caption{\label{fig2} \textbf{Spin-valley locking in Co$_{1/3}$TaS$_2$.} {\textbf{a and b,}} Side view (a) and top view (b) of the crystal structure of Co$_{1/3}$TaS$_2$. {\textbf{c,}} Fermi surface mapping of Co$_{0.33}$TaS$_2$ measured at 55\,eV (left) and the corresponding spin-polarized Fermi surface contours (right). The dashed gray lines mark the Brillouin zone of Co$_{0.33}$TaS$_2$. {\textbf{d}} Spin-integrated band structures of Co$_{0.33}$TaS$_2$ measure at 55\,eV along momentum cut A, B, and C, as marked in (c). {\textbf{e,}} Spin-resolved EDCs (upper panels) and corresponding spin polarizations (lower panels) taken at locations 1-7, as marked in (c,\,d).
}
\end{figure*}

To examine how Co intercalation modifies the spin-valley physics, we performed spin-resolved ARPES measurements on Co$_{1/3}$TaS$_2$. The side view of the crystal structure (Fig.~\ref{fig2}a) illustrates that magnetic Co atoms occupy the van der Waals gaps between adjacent 1$\it{H}$-TaS$_2$ layers, consequently breaking global inversion symmetry\,\cite{SParkin1983PBrown}. The top view (Fig.~\ref{fig2}b) further shows that the Co atoms form a $\sqrt{3}\times\sqrt{3}$ triangular lattice, rotated by 30\,$^{\circ}$ with respect to the 1$\times$1 TaS$_2$ unit cell. The Fermi surface map of Co$_{0.33}$TaS$_2$ presented in Fig.~\ref{fig2}c reveals electron-doped spin-split bands ($\alpha$ and $\beta$) along with an additional triangular electron pocket ($\gamma$) arising from the intercalated Co atoms, as discussed in Ref.\,\cite{HLuo2025ALanzara}. The corresponding Fermi surface contours are plotted on the right half of Fig.~\ref{fig2}c. Band structures measured along three representative momentum cuts (A-C in Fig.~\ref{fig2}c) are shown in Fig.~\ref{fig2}d. These data clearly demonstrate the electron-doped nature of the $\alpha$ and $\beta$ bands, as evidenced by the downward shift of their band minima toward higher binding energy relative to pristine 2$\it{H}$-TaS$_2$ (Fig.~\ref{fig1}d), as well as the electron-like dispersion of the Co-derived $\gamma$ band. Moreover, a degeneracy of the $\alpha$ and $\beta$ bands appears near $\bar\Gamma_0$ at photon energy $h\nu$\,=\,55\,eV, which is absent at $h\nu$\,=\,93\,eV\,\cite{HLuo2025ALanzara}. This difference likely arises from the fact that electronic structures probed at these two photon energies correspond to distinct $k_z$ planes, as we will discuss below.

Out-of-plane spin-resolved EDCs were measured at selected momenta to map the spin texture of Co$_{0.33}$TaS$_2$, specifically at points 1-7 (indicated by dots in Fig.~\ref{fig2}c and arrows in Fig.~\ref{fig2}d). The corresponding spectra are shown in Fig.~\ref{fig2}e: the upper panels display the spin-up and spin-down components, and the lower panels show the extracted spin polarizations. At point 1, only a single peak without detectable spin polarization is observed. This is because the $\alpha$ and $\beta$ bands remain nearly degenerate near $\bar\Gamma_0$; as a result, their strongly overlapping contributions cancel any band-resolved spin polarization, yielding essentially zero net signal. In contrast, spectra at points 4-7 show two well-resolved peaks with opposite dominant spin components, indicating that the $\alpha$ and $\beta$ bands are spin-split near the $\rm\bar{K}_0$/$\rm\bar{K}'_0$ points, and that their spin polarizations reverse upon $\bar\Gamma_0$-$\rm\bar{M}_0$ reflection. For points 2 and 3, a single peak from the Co-derived $\gamma$ band\,\cite{HLuo2025ALanzara} appears near the Fermi level, with no detectable out-of-plane spin polarization within our experimental resolution. Additional spin-resolved data at other momenta across the Fermi surface are provided in Supplementary Fig.~S1. The extracted spin texture is shown in the right half of Fig.~\ref{fig2}c. Blue and red curves trace Fermi surface segments with dominant spin-up and spin-down components, respectively, while gray curves mark regions with negligible spin polarization. These results confirm that the TaS$_2$-derived $\alpha$ and $\beta$ bands retain spin-valley locking behavior, whereas the Co-derived $\gamma$ band is spin-degenerate.
\vspace{4mm}

\begin{figure*}[!htb]
\begin{center}
\includegraphics[width=1.7\columnwidth,angle=0]{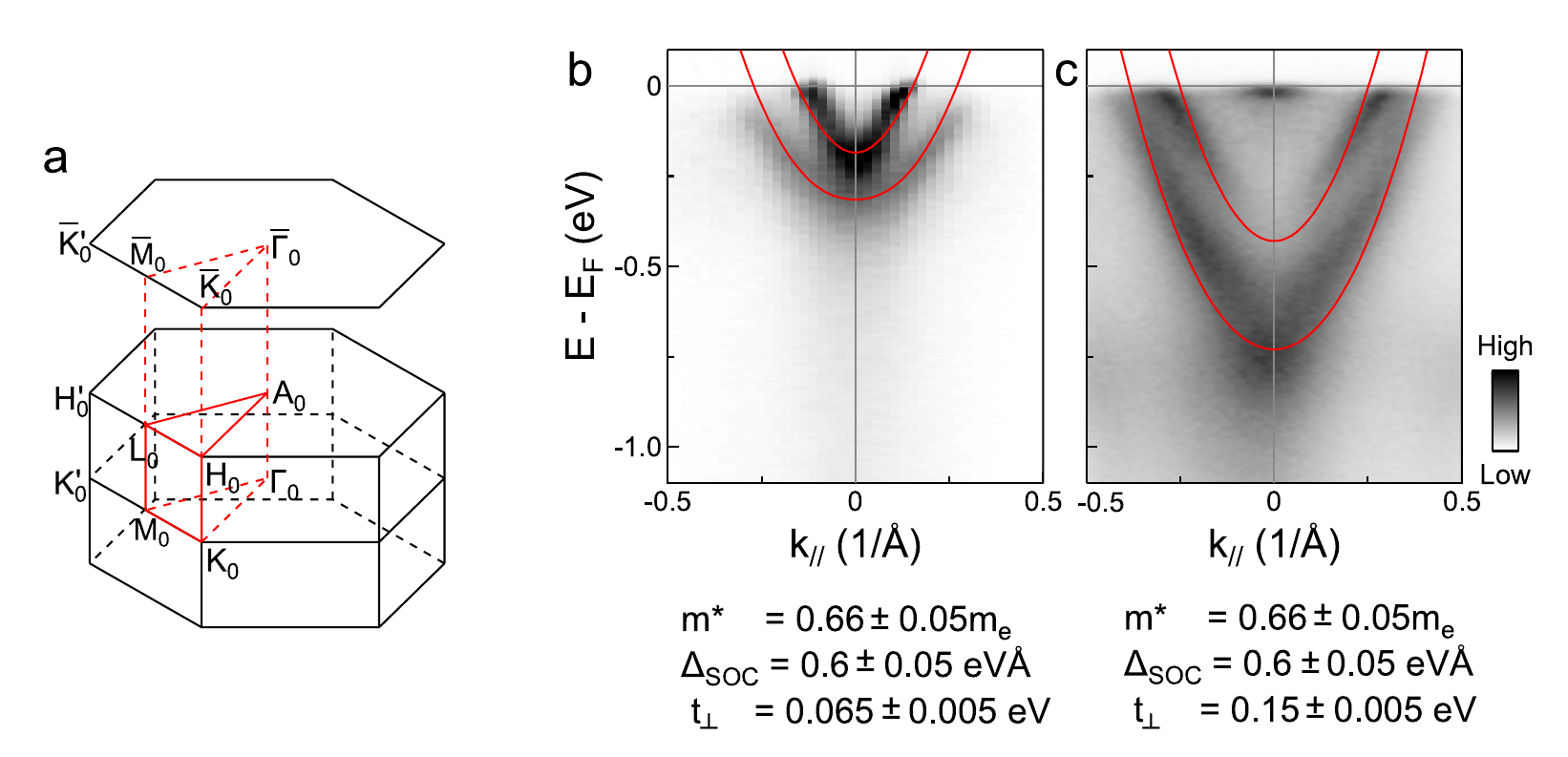}
\end{center}
\caption{\label{fig3} \textbf{Enhanced interlayer coupling mediated by Co intercalation in Co$_{1/3}$TaS$_{2}$.} {\textbf{a,}} Three-dimensional Brillouin zone of 2$\it{H}$-TaS$_2$ with high-symmetry points, and its projection onto the two-dimensional plane. {\textbf{b-c,}} Band structures of 2H-TaS$_2$\,(b) and Co$_{0.33}$TaS$_{2}$\,(c) along the $\rm\bar{K}_0$-$\rm\bar{M}_0$-$\rm\bar{K}_0'$ direction. Solid lines represent calculated bands based on a simple tight-binding model. The extracted values of the effective mass\,(m*), spin-orbit coupling strength\,($\rm\Delta_{SOC}$), and interlayer coupling strength\,($t_{\perp}$) are listed below each spectrum.
}
\end{figure*}

To elucidate the role of Co atoms in shaping the spin texture of Co$_{1/3}$TaS$_2$, we consider two possible contributions: the introduction of magnetism and the modification of interlayer coupling. Regarding magnetism, the intercalated Co atoms form a long-range antiferromagnetic order with a weak out-of-plane ferromagnetism\,\cite{HTakagi2023SSeki,PPark2023JPark,PPark2024JPark}. In principle, exchange coupling between the localized Co moments and itinerant electrons in the TMD layer could modify the spin-valley splitting, leading to an asymmetry in the $\alpha$/$\beta$ band splitting between the $\rm\bar{K}_0$ and $\rm\bar{K}'_0$ valleys\,\cite{BEdwards2023PKing}. However, no discernible asymmetry is observed between these valleys (Fig.~\ref{fig2}), suggesting that the weak ferromagnetism in Co$_{1/3}$TaS$_2$ alone is insufficient to effectively alter the valley-spin splitting.

To examine the effect of Co intercalation on interlayer coupling, we compare the band structures along the $\rm\bar{K}_0$-$\rm\bar{M}_0$-$\rm\bar{K}'_0$ direction (Fig.~\ref{fig3}a) for 2$\it{H}$-TaS$_2$ and Co$_{0.33}$TaS$_2$ (Fig.~\ref{fig3}b-c). In 2\textit{H}-TaS$_2$, interlayer coupling between adjacent 1\textit{H} layers induces a band splitting that is maximized at $\rm{M}_0$ and vanishes at $\rm{L}_0$, where glide-mirror symmetry protects the degeneracy\,\cite{LBawden2016PKing,DDentelski2021JRuhman}. Based on the inner potential reported in Ref.\,\cite{AAlmoalem2024AKanigel}, we chose photon energies of $h\nu$ = 98\,eV and 80\,eV to probe 2\textit{H}-TaS$_2$ and Co$_{0.33}$TaS$_2$, respectively, whose $c$-axis lattice constants are 12.1\,\AA~\cite{SNagata1992KTsutsumi} and 11.9\,\AA~\cite{PPark2024JPark}. Under these conditions, both measurements probe approximately the $k_z\,=\,0$ plane, where the interlayer-coupling-induced band splitting is expected to be maximal along $\rm{M}_0$-$\rm{L}_0$.

A simple and effective model consisting of two coupled Ising layers can be used to extract the interlayer coupling strength from our data:
\begin{equation}
H = \frac{\hbar^2}{2m^*} \sigma_0 \tau_0 k^2 + \Delta_{\mathit{SOC}}\,\sigma_z \tau_z k + \sigma_0 \tau_x t_{\perp}
\end{equation}
Here, $\sigma$ and $\tau$ denote Pauli matrices for spin and layer degrees of freedom, respectively; $\Delta_{\mathit{SOC}}$ represents the Ising SOC strength, and $t_{\perp}$ denotes the interlayer coupling. For a meaningful comparison of the interlayer coupling, the effective mass and Ising SOC are set to be equal in both materials and are estimated to be 0.66\,$\pm$\,0.05\,m$\rm{_e}$ and 0.6\,$\pm$\,0.05\,eV$\rm\AA$, respectively. Fitting the model to our data yields an $t_{\perp}$\,=\,65\,$\pm$\,5\,meV for 2$\it{H}$-TaS$_2$ and 150\,$\pm$\,5\,meV for Co$_{0.33}$TaS$_2$, with the former consistent with previous reports\,\cite{SBarrera2018BHunt,AAlmoalem2024AKanigel}. The Co-intercalated compound thus exhibits an interlayer coupling about 2.5 times larger, even though the interlayer spacing shrinks only slightly (from 6.05\,\AA~in 2$\it{H}$-TaS$_2$ to 5.95\,\AA~in Co$_{0.33}$TaS$_2$), which alone cannot account for such a substantial enhancement. This indicates that Co intercalants actively mediate interlayer hopping and strongly boost interlayer hybridization. A previously proposed hybridization-amplification mechanism in the sister compound Co$_{1/3}$NbS$_2$\,\cite{PPopcevic2022ETutis,PPopcevic2023ETutis} may offer a explanation here: the interlayer hopping between Ta 5$d$ orbitals is significantly reinforced by their hybridization with the intercalated Co 3$d{z^2}$ orbitals. This behavior contrasts sharply with 2$\it{H}$-TaS$_2$ systems intercalated with non-magnetic species such as SnS, 1$\it{T}$-TaS$_2$, or chiral molecules, where interlayer interaction is markedly suppressed\,\cite{Zli2024JLu,AAlmoalem2024AKanigel,ZWan2024XDuan}. These findings highlight the unique role of Co intercalants in promoting hybridization-amplified interlayer coupling in layered TMDs.
\vspace{4mm}


\begin{figure*}[!htb]
\begin{center}
\includegraphics[width=2\columnwidth,angle=0]{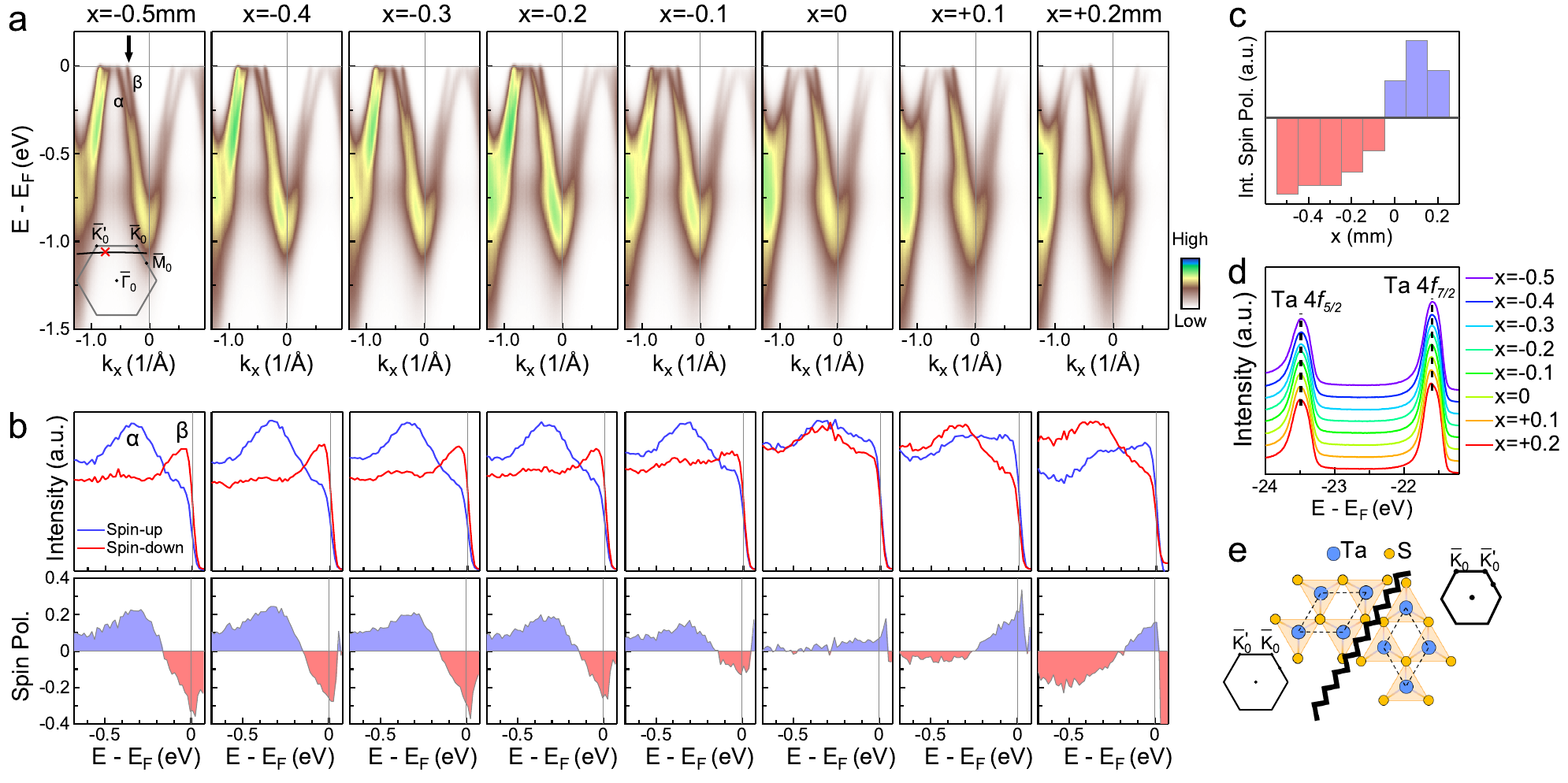}
\end{center}
\caption{\label{fig4} \textbf{Spatially reversed spin polarization on the surface of a Co$_{0.33}$TaS$_{2}$ sample.} {\textbf{a,}} Band structures of Co$_{0.33}$TaS$_{2}$ taken along the momenta path indicated by the black curve in the inset of the leftmost panel. These spectra were acquired along a horizontal line across the sample with a step size of 0.1\,mm. The coordinate $x$ of the measurement positions are labeled above each spectrum. {\textbf{b,}} Corresponding out-of-plane spin-resolved EDCs (upper panels) and spin polarizations (lower panels) taken at all positions shown in (a), at the momentum location marked by the black arrow and red cross in (a). {\textbf{c,}} Integrated spin polarization of the $\beta$ band as a function of lateral sample position $x$. {\textbf{d,}} Core-level spectra collected from all positions between  x\,=\,$-$\,0.5\,mm and  x\,=\,+\,0.2\,mm. The dashed lines indicate the peak positions of the Ta 4$f_{5/2}$ and Ta 4$f_{7/2}$ orbitals. {\textbf{e,}} Schematic illustration of two adjacent layers or two rotational domains with 180$^\circ$ in-plane rotation, along with their corresponding Brillouin zones.
}
\end{figure*}

Using the $c$-axis lattice constants together with our experimentally determined $\Delta_{\mathit{SOC}}$ and $t_{\perp}$ in the established expressions ($P\propto\frac{1}{\sqrt{1 + \left(t_{\perp}/\Delta_{\rm{SOC}}\right)^2}}$ and $P\propto\frac{1}{1 + e^{-\Delta L / L}}$), we estimate that the measured spin polarization from the topmost layer of 2$\it{H}$-TaS$_2$ and Co$_{0.33}$TaS$_2$ should be $\sim$\,70\% and $\sim$\,68\% of that of an isolated 1$\it{H}$-TaS$_2$ monolayer. These similar values are consistent with the comparable spin polarizations observed experimentally in Fig.\ref{fig1} and Fig.\ref{fig2}. However, spatially resolved measurements reveal pronounced variations in spin polarization across the sample surface. Further analysis attributes this to the presence of two distinct surface domains exhibiting opposite spin polarization (Fig.~\ref{fig4}). This extrinsic effect introduced considerable uncertainty in determining the intrinsic spin texture.

Figure~\ref{fig4}a shows the band structure of Co$_{0.33}$TaS$_{2}$ along the momentum path marked in the inset, capturing the $\alpha$ and $\beta$ bands near $\rm\bar{K}_0$ and $\rm\bar{K}'_0$. We then moved the sample laterally, acquiring eight spectra from  x\,=\,$-$\,0.5\,mm to +0.2\,mm in 0.1\,mm steps. Although some redistribution of spectral weight is seen, the overall band dispersions remain essentially unchanged. Spin-resolved EDCs at each position, taken at the momenta indicated by the black arrow and red cross in Fig.~\ref{fig4}a, are displayed in Fig.~\ref{fig4}b. In the leftmost panels (at x\,=\,$-$\,0.5\,mm), the $\beta$ band exhibits a dominant spin-down component, whereas the $\alpha$ band is predominantly spin-up. As x is increased, the spin polarization gradually weakens and eventually reverses for both bands. For clarity, Fig.~\ref{fig4}c plots the integrated spin polarization of the $\beta$ band versus positions, highlighting the spatial reversal of spin polarization.

To investigate the origin of this spatially reversed spin texture, we measured Ta $4f_{5/2}$ and $4f_{7/2}$ core-level spectra (Fig.~\ref{fig4}d), which exhibit no discernible shift within experimental resolution. The unchanged band structure and core-levels contrast with cases of distinct surface terminations (e.g., Co-termination vs TaS$_2$-termination), where a clear difference in chemical potential is observed\,\cite{BEdwards2023PKing,HLuo2025ALanzara}. A plausible explanation lies in the presence of two in-plane domains rotated by 180$^\circ$, as illustrated in Fig.~\ref{fig4}e. Such neighboring domains can arise either from orientational variants formed during growth, as reported for epitaxially grown monolayer TMDs on substrates\,\cite{YMa2017MBatzill}, or from step-like surface terraces that expose both the A and B TaS$_2$ layers after cleaving. Given that mirror-twin grain boundaries are rarely reported in bulk crystals, the terrace scenario therefore appears the more likely origin here, and this interpretation is further suggested by the ubiquitous odd-layer step terraces seen in the surface topography measurements (see Supplementary Fig.~S5). These distinct domains exhibit identical band dispersions but opposite spin polarization due to the interchange of $\rm\bar{K}$ and $\rm\bar{K}'$ valleys in momentum space. This near-zero polarization around $x\,=\,0$\,mm is therefore naturally explained by averaging over two domain types of comparable areas. Notably, because the ARPES probing spot ($\sim$\,100\,$\mu$m) is many orders of magnitude larger than the lattice constants and necessarily covers a large number of unit cells, we cannot exclude the possibility that the data collected near x\,=\,$-$\,0.5\,mm and x\,=\,$+$\,0.2\,mm also contain contributions from both domain types. However, the large and spatially uniform spin polarization there implies that one domain type dominates within the probed region, even though a minority domain may still be present. Similar domain-related spin reversal is frequently observed in other Co$_{1/3}$TaS$_2$ samples (see Supplementary Fig.~S2, S3), and also in 2$\it{H}$-TaS$_2$ (see Supplementary Fig.~S4). Notably, the stronger surface corrugation in cleaved 2$\it{H}$-TaS$_2$, makes it much less likely to find two neighboring regions with equally sharp, nearly identical bands but opposite spin polarization, which may explain why such spatially reversed spin polarizations have not been reported previously in 2$\it{H}$-TaS$_2$\,\cite{AAlmoalem2024AKanigel} and other 2$\it{H}$-phase TMDs\,\cite{RSuzuki2014YIwasa}. These results underscore the critical importance of spatially resolved measurements in layered TMD systems: only data taken from regions dominated by a single domain can reliably reveal the intrinsic spin polarization. More broadly, they highlight a practical limitation for spintronic devices, where surface domains in 2$\it{H}$-TMDs may can cause spin-signal cancellation between oppositely polarized $\rm\bar{K}_0$ and $\rm\bar{K}'_0$ valleys, posing a major obstacle to generating usable spin currents at device-relevant scales\,\cite{OClark2022JBarriga}.
\vspace{4mm}

\begin{figure*}[!htb]
\begin{center}
\includegraphics[width=2\columnwidth,angle=0]{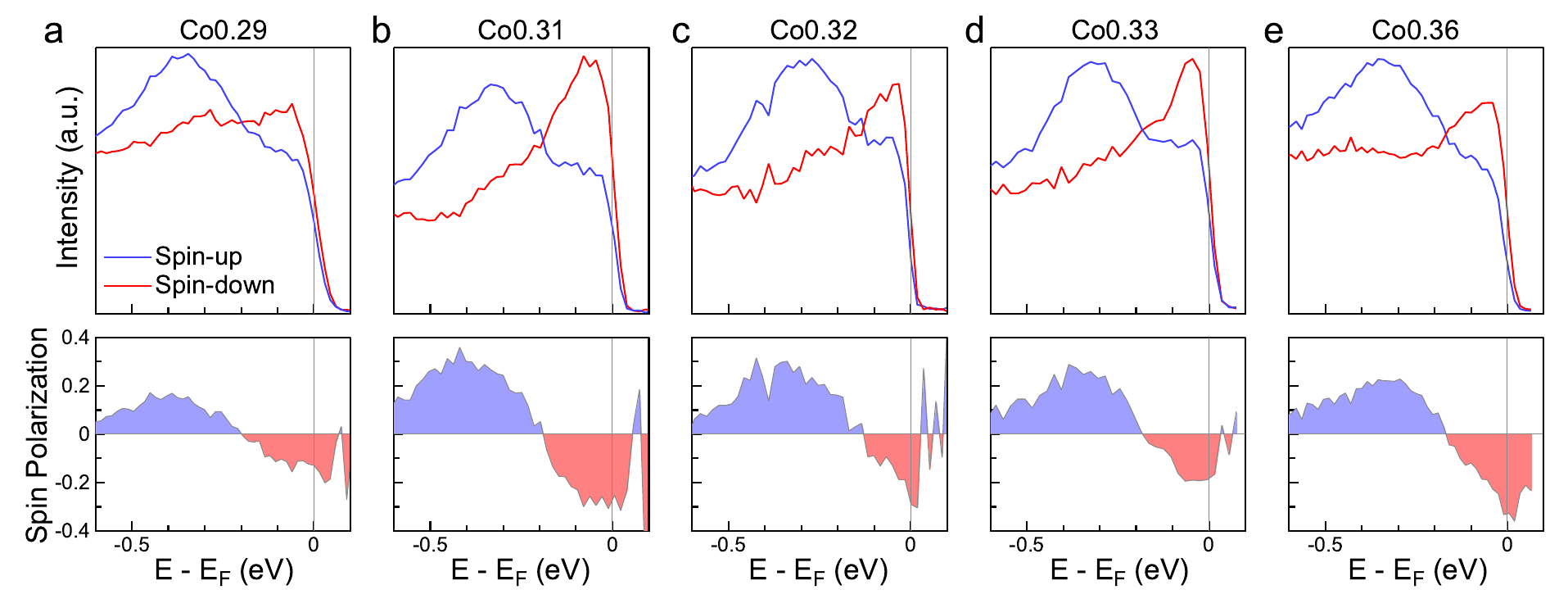}
\end{center}
\caption{\label{fig5} \textbf{Doping dependence of spin-resolved electronic structure in Co$_x$TaS$_2$.} {\textbf{a-e,}} Spin-resolved EDCs measured on samples with Co compositions of $x$ = 0.29\,(a), 0.31\,(b), 0.32\,(c), 0.33\,(d), and 0.36\,(e) are shown in upper panels. Corresponding spin polarizations are displayed in lower panels. All spectra were acquired at the same momentum location as in Fig.~\ref{fig4}.
}
\end{figure*}

To further examine the robustness and doping dependence of the surface spin polarization in Co$_x$TaS$_2$, we performed spin-resolved measurements on samples with Co compositions ranging from $x = 0.29$ to $0.36$ (Fig.~\ref{fig5}). For each doping level, multiple crystals were measured, and we report data from locations where spatial scans revealed no significant variations in spin polarization between nearby points. This invariance suggests a single-domain dominated region.

Across the doping series, all samples exhibit maximum spin polarizations of roughly 20$\%$$\sim$30$\%$, comparable to pristine 2$H$-TaS$_2$. Overall, we do not observe any abrupt change in spin polarization across the magnetic phase boundary ($x \approx 0.33$). This supports the interpretation that the spin polarization primarily originates from spin-valley locking of parent 2$\it{H}$-TaS$_2$, which is relatively insensitive to Co doping within the examined range.

\vspace{4mm}

Our study reveals that spin-valley locking is preserved at the surfaces of both pristine and magnetically intercalated 2$\it{H}$-TaS$_2$. This persistence complicates efforts to isolate spin signatures arising specifically from altermagnetism in intercalated 2$\it{H}$-TMD systems\,\cite{SFender2025KBediako}. Meanwhile, the observed spin polarization is intricately influenced by the underlying domain structure. The discovery of domain-dependent spin inversion, most likely originating from step-like surface terraces, underscores a critical yet previously underappreciated source of ambiguity in spin-resolved measurements on TMDs. While Co intercalation introduces antiferromagnetism and enhances interlayer hybridization, it does not significantly alter the intrinsic surface spin texture. These findings highlight the subtle interplay between symmetry, interlayer coupling, and domain structure in shaping the spin-valley landscape. Looking ahead, combining spin-ARPES with scanning-probe techniques such as scanning tunnelling microscopy or atomic force microscopy on the same cleaved surfaces will be highly valuable for directly correlating domain structure with the measured spin polarization, thereby helping to unambiguously interpret the resulting data. In parallel, precise control over surface domains\,--\,via selective cleaving, epitaxial growth engineering, or domain egineering\,--\,will be essential for reliably manipulating spin textures in TMD systems. This work establishes a foundation for exploring spin-dependent quantum phenomena in layered and intercalated materials with micron-scale spatial resolution.
\vspace{6mm} 

\noindent {\bf Supporting Information}

\vspace{2mm}
Methods for growth and characterization of single crystals, high-resolution ARPES measurements, analysis of the spin-ARPES data; additional data on spin-valley locking in 2H-TaS$_2$ and Co$_{0.32}$TaS$_2$, spatially reversed spin polarization on the surface of Co$_x$TaS$_2$ and 2H-TaS$_2$, and AFM characterization of cleaved Co$_{0.33}$TaS$_2$ and 2H-TaS$_2$.

\vspace{3mm}

\noindent {\bf Acknowledgement}\\
This work was primarily supported by the U.S. Department of Energy, Office of Science, Office of Basic Energy Sciences, Materials Sciences and Engineering Division, under contract No. DE-AC02-05CH11231 (Quantum Materials Program KC2202). This research used resources of the Advanced Light Source, a US DOE Office of Science User Facility under Contract No. DE-AC02-05CH11231.
\vspace{2mm}

\noindent {\bf Author Contributions}\\
A.L. and H.-L.L. conceived the project. H.-L.L. carried out the ARPES measurements with assistance from M.H., H.J., L.M., C.J., A.B., and A.F. H.-L.L. performed the data analysis and calculations. J.R., C.X., and J.G.A. synthesized the single crystals. P.T.M. and J.G.A. performed the AFM measurements. D.H.L. contributed to theoretical analysis. The manuscript was written by H.-L.L. and A.L., with input from all authors. All authors contributed to discussions and provided feedback on the manuscript.
\vspace{2mm}

\noindent {\bf Competing interests}\\
The authors declare no competing interests.
\vspace{2mm}

\noindent {\bf Data and materials availability}\\
All data are available in the manuscript.
\vspace{2mm}

\noindent {\bf References}\\

\end{document}